\documentstyle[12pt]{article}

\newcommand{\e}[1]{\label{eq:#1}}
\newcommand{\ee}[1]{(\ref{eq:#1})}

\newcommand{\eq}{\begin{equation}}
\newcommand{\eqe}{\end{equation}}
\newcommand{\eqa}{\begin{eqnarray}}
\newcommand{\eqae}{\end{eqnarray}}

\newcommand{\del}{\partial}

\newcommand{\bra}[1]{\mbox{$\langle #1 $}}
\newcommand{\ket}[1]{\mbox{$| #1 \rangle$}}

\begin{document}

\pagestyle{empty}
\hfill{NSF-ITP-94-19}

\hfill{hep-th/9402156}

\vspace{18pt}

\begin{center}
{\large \bf Gravitational Scattering in the $c=1$ Matrix Model}

\vspace{16pt}
Makoto Natsuume and Joseph Polchinski

\vspace{16pt}
\sl

Institute for Theoretical Physics \\ University of California \\
Santa Barbara, California 93106-4030

\rm

\vspace{12pt}
{\bf ABSTRACT}

\end{center}

\begin{minipage}{4.8in}
The $c=1$ matrix model is equivalent to $1+1$ dimensional string theory.
However,
the tachyon self-interaction in the former is local, while in the
latter it is nonlocal due to the gravitational, dilaton and higher string
fields.
By studying scattering of classical pulses we show that the appropriate
nonlocal field redefinition converts the local matrix model interaction into
the expected string form.  In particular, we see how the asymptotic behavior of
the gravitational field appears in the scattering.

\end{minipage}

\vfill

\pagebreak
\pagestyle{plain}
\setcounter{page}{1}
\setcounter{section}{-1}
\baselineskip=16pt
\section{A Digression}

The exact solution of low dimensional string theory by means of matrix models
was a remarkable discovery\cite{mmrefs,mmreviews}.  In spite of the
substantial effort in this area, one must feel that the physical content of the
solution has not been fully developed.  In this paper we report on a further
step in this direction, after a brief discussion of some general issues.

It is sometimes said that little has been learned from the
matrix models.  This is not true.  Matrix
models have taught us a vital lesson: that the ``Theory of Everything''
is {\it not} string theory.  Let us elaborate, focussing first on the closely
related string theory of two-dimensional $U(N)$ gauge theory\cite{GrossT}.
Canonically quantized on a circle, a typical invariant state is
\eq
{\rm tr}(U)^{n_1} {\rm tr}(U^2)^{n_2} \ldots {\rm tr}(U^m)^{n_m},
\e{traces}
\eqe
where $U$ is the holonomy around the circle.  For convenience we focus on one
chiral sector---that is, positive powers of $U$.  The trace
${\rm tr}(U^k)$ can be associated with a string that winds $k$ times
around the circle\cite{MinPol,Douglas}.  In particular, one can introduce
creation and annihilation operators for the $k$-times wound string,
\eq
[a_k, a_l^\dagger] = k \delta_{k,l}. \e{algebra}
\eqe
The state~\ee{traces} is then proportional to
\eq
{a_1^\dagger}^{n_1} {a_2^\dagger}^{n_2} \ldots {a_m^\dagger}^{n_m} \ket{0}
\e{RL}
\eqe
where $\ket{0}$ is the constant wavefunction.  The Hamiltonian can be written
as a string tension plus a splitting-joining interaction,
\eq
H = \frac{g^2 L}{2} \sum_{k=1}^\infty a^\dagger_k a_k
+ \frac{g^2 L}{2 N} \sum_{k,k'=1}^\infty ( a^\dagger_{k+ k'} a_k a_{k'}
+ {\rm h.c.}),
\eqe
as well as a contact
(zero-size handle) term depending on the $U(1)$ factor.

As long as the total number of string windings $\sum_{k=1}^\infty k n_k$ is
less than $N$, the states~\ee{traces} are independent and in fact orthogonal
under the group integration, as implied by the representation~\ee{RL} and the
algebra~\ee{algebra}.  But for $N$ or more windings this fails: for example
${\rm tr}(U^N)$ can be expanded in terms of lower traces.  So while the stringy
Hamiltonian correctly reproduces the perturbation series in $1/N$, it fails
non-perturbatively.  The point is not merely that there are non-analytic terms
in the $1/N$ expansion, but the stringy description itself, the enumeration of
states, is breaking down.\footnote{This point arose in discussions with M.
Douglas, A. Strominger and M. Stone at the ITP Workshop on Nonperturbative
String Theory. The need to supplement the string description with a projection
has also been discussed recently by Taylor (seminar at UCSB).}  In this case a
better description is known---the theory can be put in fermionic form and this
description is  exact\cite{MinPol,Douglas}.
The breakdown of the string picture has a simple interpretation in the
fermionic
language.  The number of windings corresponds to the total number of levels by
which the fermions are promoted from the ground state.  The bosonic description
does not know that the Fermi sea has both an upper and a lower edge (with a
total of $N$ filled levels) and for $N$ or more windings it includes states
where a fermion is promoted from below the lower edge to above the upper---but
the former state is actually empty to start with.

Exactly the same issue arises in the $c=1$ matrix model.  The fermionic
description is well-defined.  The bosonic (string) description is valid near
one edge of the Fermi surface but breaks down when both the upper and lower
edge become involved.  This is nonperturbative in the string coupling,
occuring when the density of string is of order $1/g_{\rm s}$.  Again, this
is not like field theory, where the perturbation series is asymptotic but the
theory is in principle exact---here the string theory itself is only an
asymptotic description and new variables are needed.

It could be that this is special to the matrix model and does not apply to
higher-dimensional strings, but there are several signs that it is general.
One is the non-field theoretic $e^{-O(1/g_{\rm s})}$ nonperturbative
behavior\cite{Shenker}.  Another is the unwieldiness of string field
theory---the need to correct the covariant closed string theory at each order
of perturbation theory\cite{Zwiebach}, and the fact that the related
space of all two-dimensional field theories does not seem to have a natural
definition.

So we are proposing that not only is string perturbation theory merely
asymptotic, but that string theory itself only generates the asymptotics of the
``Theory of Everything.''  Finding the correct description is vital, both
because the $e^{-O(1/g_{\rm s})}$ effects are apt to be numerically as or more
important than the familiar $e^{-O(1/g_{\rm s}^2)}$ effects, and because we
might hope that it will bring in new concepts that are essential to
understanding such issues as the physics of the vacuum.

\section{Introduction}

We now return to a narrower issue---finding spacetime gravitational dynamics
in the matrix model.
The dilaton-graviton sector of two-dimensional string theory should have
interesting dynamics, including black holes\cite{Witt,MSW}.  Further, this
string theory is exactly solvable through the $c=1$ matrix
model\cite{c1refs}.  One would like to make use of this solution to
address basic questions, including the effect of string theory on spacetime
singularities and the full quantum evolution of the black hole.  But while
various proposals have been made, it is not clear how the black hole
background is described in the matrix model.  It should be possible to
study the same processes as in dilaton gravity\cite{CGHS} and its
generalizations, where pulses containing energy and information are sent
toward the strong coupling region and a black hole forms and then
evaporates.  Or, if this process does not occur in the string theory, one would
like a clear understanding of why this is the case.

There is an argument which would appear to indicate that gravitational effects
are for some reason absent in the matrix model.  Imagine sending two matter
(tachyon) pulses toward the strong coupling region, one after the other, as
shown in figure~1.  The first pulse carries energy and so will produce a
gravitational field; the second pulse will then have some amplitude to
back-scatter off this field. But in the matrix model, these pulses are
packets of non-interacting fermions which travel freely in the inverted
harmonic oscillator potential\cite{meclass}.  The first packet thus
does not affect the motion of the second.

The resolution of this paradox is in principle known, though it has not been
developed in this time-dependent context.  The back-scattering process is
``bulk" scattering, which is indeed absent in the matrix model\cite{s1c1}.
However, the string S-matrix differs by a certain wavefunction
renormalization and has nonzero bulk
scattering\cite{PolyX,dFK,MYL}.  The renormalization, although
linear in the fields and merely a phase for real momenta, is able to convert
an interacting theory into a non-interacting one because the kinematics
restricts the scattering to particular points in the complex momentum plane
where the renormalization factor vanishes.

It is sometimes stated that this wavefunction renormalization, being a phase,
does not affect probabilities and so can be ignored. But an energy-dependent
phase produces a time delay, and we are specifically interested in
time-dependent processes.\footnote{Put differently, the renormalization is a
pure phase only in a particular basis, and the states of interest to us will
necessarily be superpositions of these basis elements.} Indeed we will see
that the renormalization plays an essential role.

In this paper we consider only the classical scattering of pulses that are
not too large, in that the Fermi surface remains single valued and does not
pass over the potential barrier.  In a sense all of our results are then
obvious a priori, from refs.~\cite{PolyX,dFK,MYL}.  But given the
confusion in this subject, it is worth working out in detail this point of
contact between the matrix model and the continuum string theory.  The
calculation is slightly convoluted, and is a necessary preliminary
to studying the more interesting dynamics of large pulses.

\section{Review of Matrix Model Scattering}

We first review the classical solutions to the matrix model, following
refs.~\cite{meclass,MP}.  The Hamiltonian is
\eqa
H &=& \frac{1}{2} \int_{-\infty}^\infty dx \Bigl\{ \del_x
\psi^\dagger \del_x \psi - x^2 \psi^\dagger \psi \Bigr\}  \nonumber\\
&=& \frac{1}{2\pi} \int_{-\infty}^\infty
 dx \biggl\{ \frac{1}{6}(p_+^3 - p_-^3)
- \frac{x^2}{2} (p_+ - p_-) \biggr\}
\eqae
where $p_{\pm}$ are the upper and lower surfaces of the Fermi sea.  Our
conventions are as in ref.~\cite{meclass} with two changes.
We now
set $\alpha' = 1$ so that the matrix model embedding time coincides with
that of the continuum theory; this also simplifies most
expressions.\footnote {In ref.~\cite{meclass}, only eq.~(44) is affected by
this.} And, we now omit the factors of $g_{\rm s}$ from
the definitions~(11) of that paper in order that the Hamiltonian be independent
of $g_{\rm s}$.\footnote
{Let us also here note two misprints in ref.~\cite{meclass}.  Eq.~(25)
should read $x = - e^{-q}$.  In eq.~(34) the last $\pm$ should be
$\mp$---the corrected form is given in eq.~\ee{aseps} below, now with
$g_{\rm s}$-dependence.}  Then $g_{\rm s}$ enters only as  a
parameter in the static solution,
\eq
p_{\pm} = \pm \sqrt{x^2 - g_{\rm s}^{-1}}.
\eqe
Focusing on one side of the barrier, say $x < 0$, the theory can be written
in terms of a canonically normalized massless scalar $\overline S(q,t)$,
where $x = -e^{-q}$:
\eqa
p_{\pm}(x,t) &=& \mp x \pm \frac{1}{x} \epsilon_\pm(q,t)   \nonumber\\
\epsilon_\pm(q,t)/{\sqrt{\pi}} &=& \pm \overline\Pi(q,t) -
\del_q \overline S(q,t).
\eqae
Here we introduce a bar to distinguish
the matrix model objects here from the related string theory objects to be
discussed in the next section.  The Hamiltonian takes the form
\eq
H = \frac{1}{2} \int_{-\infty}^\infty
 dq \Bigl\{ \overline\Pi^2 + (\del_q \overline S)^2 +
e^{2q} O(\overline S^3)
\Bigr\}.
\eqe
The trilinear coupling vanishes as $e^{2q}$ in the asymptotic
region $q \to -\infty$ and $\overline  S$ can be expanded asymptotically as the
static solution plus a massless free field,
\eqa
\overline  S(q,t) &\sim& -\frac{q}{2 \sqrt{\pi} g_{\rm s}} +
\overline  S_+(t-q) + \overline  S_-(t+q) \nonumber\\
\overline S_{\pm}(t \mp q) &=&
\int_{-\infty}^\infty \frac{d\omega}{2\pi} \frac{1}{\sqrt{2}i \omega}
\overline\alpha_\pm (\omega) e^{i\omega(t \mp q)}.
\eqae
Asymptotically,
\eqa
\epsilon_\pm(t\mp q) &\sim& \frac{1}{2 g_{\rm s}}
+ \delta_{\pm}(t \mp q \pm \ln \sqrt{4g_{\rm s}})
\nonumber\\
\delta_{\pm}(t \mp q) &=& \pm
\sqrt{2\pi} \int_{-\infty}^\infty \frac{d\omega}{2\pi}\,
\overline\alpha_\pm (\omega)
(4 g_{\rm s})^{\mp i \omega/2} e^{i\omega(t \mp q)}.   \e{aseps}
\eqae

Classical solutions are described by the Fermi surface moving freely in the
inverted potential.  The outgoing Fermi surface is related to the incoming
surface in a nonlinear way through the time delay:
\eq
\epsilon_-(u) = \epsilon_+(u'), \qquad u' = u + \ln(\epsilon_-(u)/2).
\e{delay} \eqe
By changing variables $u \to u'$ and using~\ee{delay}, one finds\cite{MP}
\eq
\int_{-\infty}^\infty du\,(\epsilon_-(u))^r e^{-i\omega u}
=  2^{-i\omega} \frac{r}{r + i \omega}
\int_{-\infty}^\infty du'\,
(\epsilon_+(u'))^{r+ i \omega} e^{-i\omega u'}
\eqe
for arbitrary parameters $\omega$ and $r$.
Expanding around the static background as in~\ee{aseps}
gives
\eq
\delta_-(u) = \sum_{n=1}^\infty \frac{(2g_{\rm s})^{n-1}}{n!}
\frac{\Gamma(1 + \del_u)}{\Gamma(2-n+ \del_u)} (\delta_+(u))^n
\eqe
or
\eqa
\overline\alpha_-(\omega)
&=& -\sum_{n=1}^\infty \frac{(g_{\rm s}\sqrt{8\pi})^{n-1}}{n!}
\frac{\Gamma(1 + i\omega)}{\Gamma(2-n+i\omega)}
(4g_{\rm s})^{-i \omega} \e{mmsol}\\
&&\qquad\qquad\qquad \biggl\{ \prod_{i=1}^n \int \frac{d\omega_i}{2\pi}
\overline\alpha_+(\omega_i) \biggr\}
2\pi\delta(\omega - {\textstyle{\sum_{i=1}^n}
\omega_i}). \nonumber
\eqae

This classical result becomes a tree-level operator statement,
giving
the tree-level S-matrix.  For example, the $n \to 1$
amplitude is
\eqa
\overline {S}_{ \omega_1,\ldots,\omega_n \to \omega_{n+1} }
&\!=\!&
\bra{0}| \overline\alpha_-(-\omega_{n+1})
\overline\alpha_+(\omega_1)\ldots \overline\alpha_+(\omega_n)
\ket{0} \e{11}\\
&\!=\!& 2\pi i\delta( \omega_{n+1} - {\textstyle{\sum_{i=1}^n}
\omega_i}) \biggl\{ \prod_{i=1}^{n+1}\! \omega_i \biggr\}
\biggl( \frac{\pi}{2} \biggr)^{-i\omega_{n+1}/2}
\frac{\del^{n-2}}{\del \mu^{n-2}} \mu^{-i\omega_{n+1} -
1} \nonumber
\eqae
where $\mu^{-1} = g_{\rm s}\sqrt{8\pi}$
and $[\overline\alpha_{\pm}(\omega), \overline\alpha_{\pm}(\omega')]
= 2\pi \omega' \delta(\omega + \omega')$.

\section{Wavefunction Renormalization}

At tree level, the S-matrix of two-dimensional string theory has also been
obtained directly with continuum methods\cite{PolyX,dFK}, and differs
from the matrix model result above only by the multiplicative factor
\eq
{S}_{ \omega_1,\ldots,\omega_n \to \omega_{n+1} }
= \overline {S}_{\omega_1,\ldots,\omega_n \to \omega_{n+1}}
\biggl( \frac{\pi}{2} \biggr)^{i \omega_{n+1}/2}
\prod_{i = 1}^{n+1}
\frac{\Gamma(i \omega_i )}{\Gamma(-i \omega_i )}.
\eqe
In other words, these are equivalent under the redefinition
\eqa
\alpha_+(\omega) &=& \biggl( \frac{\pi}{2} \biggr)^{i\omega/4}
\frac{\Gamma(i \omega )}{\Gamma(-i \omega )}
\overline\alpha_+(\omega) \nonumber\\
\alpha_-(\omega) &=& \biggl( \frac{\pi}{2} \biggr)^{-i\omega/4}
\frac{\Gamma(-i \omega )}{\Gamma(i \omega )}
\overline\alpha_-(\omega).
\eqae
For real $\omega$ this is indeed just a phase.

The string theory tachyon is $ S(t,\phi) \sim  S_{+}(x^-)
+  S_{-}(x^+)$ where $\phi$ is the Liouville field, $x^{\pm} = t \pm \phi$, and
\eq
 S_{\pm}(x^\mp) =
\int_{-\infty}^\infty \frac{d\omega}{2\pi} \frac{1}{\sqrt{2} i\omega}
\alpha_\pm (\omega) e^{i\omega x^\mp}.
\eqe
The relation of this to the matrix model scalar is thus
\eq
 S_{\pm}(x^\mp) \ =\ \biggl( \frac{\pi}{2} \biggr)^{\pm \del_t / 4}
\frac{\Gamma(\pm\del_t)}{\Gamma(\mp\del_t)} \overline S_{\pm}(x^\mp)
\ =\ \biggl( \frac{\pi}{2} \biggr)^{- \del_\phi / 4}
\frac{\Gamma(-\del_\phi)}{\Gamma(\del_\phi)}
\overline S_{\pm}(x^\mp).  \e{trans}
\eqe

To describe the scattering of an incoming ($+$) string tachyon pulse, one
must (I) transform to the matrix model tachyon field via~\ee{trans},
(II) evolve the
pulse as described in the previous section, and (III) transform back.
The first and third steps can be written
\eqa
{\rm (I)}:\quad &&
\overline S_+(x^-) = \int_{-\infty}^\infty d\tau\, K(\tau)  S_+(x^- - \tau)
\nonumber\\
{\rm (III)}:\quad &&
 S_-(x^+) = \int_{-\infty}^\infty d\tau\, K(\tau) \overline S_-(x^+ -
\tau).  \e{conv}
\eqae
The same kernel appears in both transformations,
\eqa
K(\tau) &=& \int_{-\infty}^\infty \frac{d\omega}{2\pi} e^{i \omega \tau}
\biggl( \frac{\pi}{2} \biggr)^{-i\omega/4}
\frac{\Gamma(-i \omega )}{\Gamma(i \omega )}
\nonumber\\
&=& -\frac{z}{2} J_1 (z), \qquad z = 2 (2/\pi)^{1/8} e^{\tau/2}.
\eqae
This has asymptotic behaviors
\eqa
K(\tau) &\sim& -\biggl( \frac{\pi}{2} \biggr)^{-1/4} e^{\tau}, \qquad \tau \to
-\infty \nonumber\\
&\sim& \biggl( \frac{\pi}{2} \biggr)^{-1/16} \frac{e^{\tau/4}}{\sqrt{\pi}}
\cos( z + \pi/4 ), \qquad \tau \to \infty .
\eqae

Thus, if we start with a delta-function pulse at $x^- = 0$ we get a pulse
spread
out in time, with an exponential tail at negative times and a tail which grows
and oscillates more and more rapidly at late times.
The late oscillations drop out when we have
smooth wave-packets.  To see this, take $ S_+$ to be a gaussian wave packet
of width $\ell$ in time, so that
\eq
\alpha_+(\omega)\ \propto\ \omega e^{-(\omega - \omega_0)^2 \ell^2 / 2}.
\e{gwp}
\eqe
Then
\eq
\overline S_+(x^-) \ \propto\
\int_{-\infty}^\infty \frac{d\omega}{2\pi}\,
e^{i\omega x^- - (\omega - \omega_0)^2 \ell^2 / 2}
\biggl( \frac{\pi}{2} \biggr)^{-i\omega/4}
\frac{\Gamma(-i \omega )}{\Gamma(i \omega )}.
\eqe
For large negative $x^-$, one can shift the integral into the lower half-plane,
keeping it parallel to the real axis, and the integral is dominated by the
nearest feature (pole or saddle point) in this half-plane.  This is the
pole at $\omega = -i$, so
\eq
\overline S_+(x^-) \ \propto\ e^{x^-}, \qquad x^- \to -\infty  \e{preexp}
\eqe
the same as found for the delta-function.  For large positive $x^-$,
the integral
is dominated by the nearest feature in the upper half-plane.  There are no
poles, but there is a saddle near $\omega = i x^- / \ell^2$ (note that the
gaussian dominates the gamma functions at large imaginary $\omega$).
The late-time behavior is then a nearly gaussian falloff.
That is, a pulse with gaussian falloff is transformed to one which is still
localized but with an exponential spread at early times.  This
early-time exponential will play an important role.  The late oscillations
of the kernel will
play no role in the present work, though they may in more
complicated situations.

Although the individual steps are simple, the net result is more complicated
and less intuitive than the familiar matrix model evolution without the
convolutions.  In order to develop some familiarity with this, our goal in the
present paper is to see how it gives rise to the gravitational effects
discussed in the introduction.

\section{String Scattering}

The distinction between relatively slowly falling exponential wavepackets and
more rapidly falling gaussian wavepackets will be essential.  This is
because the gravitational field that we wish to detect itself falls off
exponentially, with $G_{tt}-1\ \propto\ M e^{4 \phi}$.  Thus we need much
narrower wavepackets in order to distinguish the `long-ranged' gravitational
interaction from the tachyon self-interaction, which we would expect to be
local
or at most smeared in a gaussian way.  One can then see how the convolution,
which as we have seen turns a gaussian into an exponential, can transmute the
local matrix  model interaction into a long-ranged gravitational one.

We will expand in powers of the incoming tachyon $ S_+$, as in the
solution~\ee{mmsol}.  The gravitational effect we seek appears at
third order.
To first order, the result of
convolution-evolution-convolution is
\eq
 S^{(1)}_-(x^\mp) =
- \int_{-\infty}^\infty \frac{d\omega}{2\pi}
\frac{1}{\sqrt{2} i\omega}
e^{i \omega x^\mp}
\mu^{i\omega}
\frac{\Gamma^2(-i \omega )}{\Gamma^2(i \omega )}
\alpha_+(\omega) \e{pm1}
\eqe
where again $\mu^{-1} = g_{\rm s}\sqrt{8\pi}$.
Here $\alpha_+(\omega)$ are the modes of the incoming classical pulse.
We take the incoming pulse to have a gaussian falloff
and to be centered
near $x^- = 0$,
perhaps a finite sum of terms of the form~\ee{gwp}.
The center of the outgoing pulse is then near $x^+ = 0$.  More precisely, its
parametric dependence on $g_{\rm s}$ is $x^+ \sim \ln g_{\rm s}$, because
as $g_{\rm s}$ is increased the Fermi level approaches the top of the potential
and the time delay increases.

We now wish to pull out the leading behavior of the outgoing wave at early
times, $x^+ \to -\infty$.  This is obtained by the same method as the
asymptotic behavior~\ee{preexp}, being dominated by the pole at $\omega = -i$.
Then,
\eq
S^{(1)}_-(x^+) \sim \mu
\int_{-\infty}^{\infty} du^- (x^+ - u^- - c)
e^{x^+ - u^- }  S_+(u^-) \e{first}
\eqe
with $c = 2 + 4\Gamma'(1) - \ln \mu$.
This has a simple interpretation, as we will verify by an effective Lagrangian
calculation in the next section.
Note that it is first order in the
background $\mu$. The linear term in the integrand, from
the double pole in~\ee{pm1}, comes from the linear behavior of the tachyon
background.

To second order,
convolution-evolution-convolution gives
\eqa
 S^{(2)}_-(x^+) &=& -\frac{1}{2\sqrt 2}
\int_{-\infty}^\infty \frac{d\omega\, d\omega_1}{(2\pi)^2}
\Biggl\{ \mu^{i\omega-1} e^{i \omega x^+} \\
&&\qquad\qquad
\frac{\Gamma(-i \omega)}{\Gamma(i \omega)}
\frac{\Gamma(-i \omega_1)}{\Gamma(i \omega_1)}
\frac{\Gamma(-i \omega_2)}{\Gamma(i \omega_2)}
\alpha_+(\omega_1) \alpha_+(\omega_2) \Biggr\} \nonumber
\eqae
where $\omega_1 + \omega_2 = \omega$.
The leading behavior as $x^+ \to -\infty$ is again governed by the
first pole encountered as the $\omega$ contour is shifted, parallel to the
real axis, into the lower half-plane.
Of course, as the $\omega$ contour
is shifted, $\omega_1$ and/or $\omega_2$ must also become complex.  It is most
efficient, in the sense of avoiding spurious leading
terms which actually cancel, to keep the poles in $\omega_1$ and $\omega_2$ as
far from the axis as possible by
dividing the imaginary part equally between $\omega_1$ and
$\omega_2$.  The first pole is then at $\omega = -i$.
Evaluating the residue gives
\eqa
 S^{(2)}_-(x^+) &\sim& - \frac{1}{2\sqrt{2}} e^{x^+}
\int_{-\infty}^\infty \frac{d\omega_1}{2\pi}\,
\frac{\alpha_+(\omega_1)}{\omega_1}
\frac{\alpha_+(-\omega_1-i)}{-\omega_1-i},\qquad {\rm Im}(\omega_1) =
-\frac{1}{2} \nonumber\\
&=& \frac{1}{\sqrt{2}}
\int_{-\infty}^\infty du^-\, e^{x^+ -u^-} S_+^2(u^-). \e{2otb}
\eqae
Note that because of the gaussian falloff of $ S_+$, its Fourier transform
$\alpha_+(\omega)/\omega$ is well-defined and analytic for all complex
$\omega$; in particular the position of the $\omega_1$ contour doesn't matter
in
the final step.  This is bulk scattering of two incoming tachyons into one
outgoing.  We will verify that this can be obtained from an effective
Lagrangian
in the next section, but the main features are easily understood.  The
spacetime
dependence follows from the position dependence of the coupling---the
outgoing ray of
fixed $t + \phi = x^+$ meets the incoming ray of fixed $t - \phi = u^-$ at
$2\phi_0 = x^+ - u^-$, at which point the coupling constant is $e^{2\phi_0}$.
Also, the amplitude is zeroth
order in the background $\mu$; scatterings involving the background
would involve more interactions and so are subleading as $x^+ \to -\infty$.

To third order in the incoming field,
\eqa
 S^{(3)}_-(x^+) &=& \frac{1}{6 \sqrt{2}}
\int_{-\infty}^\infty \frac{d\omega\, d\omega_1\, d\omega_2}{(2\pi)^3}
\Biggl\{ e^{i \omega x^+}(1 - i\omega) \mu^{i\omega - 2}
\\ &&
\frac{\Gamma(-i \omega)}{\Gamma(i \omega)}
\frac{\Gamma(-i \omega_1)}{\Gamma(i \omega_1)}
\frac{\Gamma(-i \omega_2)}{\Gamma(i \omega_2)}
\frac{\Gamma(-i\omega_3)}{\Gamma(i \omega_3)}
\alpha_+(\omega_1)  \alpha_+(\omega_2) \alpha_+(\omega_3)  \Biggr\} \nonumber
\eqae
with $\omega_1 + \omega_2 + \omega_3 = \omega$.
Again the leading behavior as $x^+ \to -\infty$ is given by the first pole
encountered in the lower $\omega$ plane, and again it is efficient to divide
the imaginary part equally among $\omega_1$, $\omega_2$ and $\omega_3$.  The
first pole is then at $\omega = -2i$, giving
\eqa
&&\!\!\!\!
  S^{(3)}_-(x^+) \sim - \frac{1}{12\sqrt{2}} e^{2 x^+}
\int_{-\infty}^\infty \frac{d\omega_1\, d\omega_2}{(2\pi)^2} \Biggr\{
\frac{\Gamma(-i \omega_1)}{\Gamma(i \omega_1)}
\frac{\Gamma(-i \omega_2)}{\Gamma(i \omega_2)}
 \e{cubic}\\
&&\qquad\qquad\qquad
\frac{\Gamma(i \omega_1 + i \omega_2 - 2 )}
{\Gamma(-i \omega_1 - i \omega_2+ 2 )}
\alpha_+(-i \omega_1) \alpha_+(-i \omega_2) \alpha_+(i \omega_1 + i \omega_2
- 2 ) \Biggr\},
\nonumber\\
&&\qquad\qquad\qquad {\rm Im}(\omega_1) = {\rm Im}(\omega_2) = -\frac{2}{3}.
\nonumber
\eqae
This represents bulk scattering of three incoming tachyons into one outgoing.

To identify the long-ranged gravitational interaction we now take the incoming
field to be a sum of two gaussian pulses, the first centered at $x^- = 0$
and the second at $x^- = T$.  That is,
\eq
\alpha_+(\omega) = f_{1+}(\omega) + e^{-i\omega T}f_{2+}(\omega),
\eqe
where $\omega f_{1+}$ and $\omega f_{2+}$ are both real gaussians as
in~\ee{gwp}. The derivation of eq.~\ee{cubic} still goes through, and now we
can
extract the leading $T$-dependence as we did for $x^+$ before,
thus distinguishing
the exponential gravitational interaction from the gaussian local interactions.
The gravitational field of the first pulse is second order in $f_{1+}$, and
we wish to identify the linear scattering of the second pulse in this field,
so the relevant terms from the third-order solution~\ee{cubic} are of the form
\eq
3 e^{-i\omega T} f_{2+}(-i \omega_1) f_{1+}(-i \omega_2) f_{1+}(i
\omega_1 + i \omega_2 - 2 ).
\eqe
The first two terms at large $T$ are from $\omega_1 = -i, -2i$, giving
\eqa
 S^{(3)}_-(x^+) &\sim&
-\frac{1}{4\sqrt{2}} e^{2x^+ - T} f_{2+}(-i)
\int_{-\infty}^{\infty} \frac{d\omega_2}{2\pi}
\frac{f_{1+}(\omega_2)}{\omega_2}
\frac{f_{1+}(-\omega_2-i)}{-\omega_2-i}  \nonumber\\
&&\qquad - \frac{1}{8\sqrt{2}} e^{2x^+ - 2T} f_{2+}(-2i)
\int_{-\infty}^{\infty} \frac{d\omega_2}{2\pi}
f_{1+}(\omega_2) f_{1+}(-\omega_2) \nonumber\\
&=& \frac{1}{2} e^{2x^+} \int_{-\infty}^\infty du^-\,e^{-u^-}
 S_{2+}(u^-) \int_{-\infty}^\infty dv^-\, e^{-v^-}  S_{1+}^2(v^-)
\nonumber\\
&&\quad - \frac{1}{2} e^{2x^+} \int_{-\infty}^\infty du^-\,e^{-2u^-}
 S_{2+}(u^-) \int_{-\infty}^\infty dv^-\, \dot S_{1+}^2(v^-).\e{finally}
\eqae
In the final expression, the first term is the scattering of pulse~2 from the
second-order tachyon background~\ee{2otb} produced by pulse~1, while the second
is the gravitational scattering we seek.  That is, these represent respectively
the exchange of a tachyon and a graviton between the two pulses.  Notice in
particular that the gravitational term depends on pulse~1 precisely through its
integrated energy flux.  Also, the scattering occurs where the incoming and
outgoing rays meet, which is again $2\phi_0 = x^+ - u^-$, and the
gravitational term is then proportional to $e^{4\phi_0}$.  Thus we reconstruct
the leading correction to the metric, $\delta G_{tt}\ \propto\ M e^{4 \phi}$.

\section{Effective Field Theory}

Let us now verify in detail that the results we have found are equivalent to
those from a tachyon-graviton-dilaton effective field theory.  The spacetime
action is
\eq
S = \frac{1}{2} \int dt\,d\phi\,\sqrt{-G} e^{-2\Phi}
\Bigl\{ a_1 [ R + 4(\nabla \Phi)^2 + 16 ] - (\nabla T)^2 + 4T^2 - 2 V(T)
\Bigr\}.  \e{stact}
\eqe
The absolute normalization, which does not enter into the classical solution,
is set by a shift of the dilaton $\Phi$.  The constant $a_1$ setting the
relative normalization of the graviton-dilaton and tachyon actions will
be determined implicitly by the definition of the tachyon field below.  The
relevant part of the
tachyon self-interaction is
\eq
V(T) = a_2 T^3 / 3.
\eqe
A local quartic interaction will not contribute to the processes we consider
because of our use of wavepackets to resolve the interactions in time.  The
cubic interaction could have been a function of the tachyon momenta, but is
known from the vertex operator calculation of the three-point amplitude to be
constant up to field redefinition; we verify this below.  Other
higher-dimension operators are expected not to affect the leading $x^+ \to
-\infty$ behavior that we consider.

The field equations are
\eqa
&& R_{\mu\nu} + 2 \nabla_\mu \nabla_\nu \Phi - a_1^{-1} \del_\mu T \del_\nu T =
0 \nonumber\\
&& R + 4 \nabla^2 \Phi - 4 (\nabla \Phi)^2 + 16 - a_1^{-1} (\nabla T)^2
+4 a_1^{-1} T^2 - \frac{2}{3} a_2 a_1^{-1} T^3 = 0 \nonumber\\
&& \nabla^2 T - 2 \nabla \Phi \nabla T + 4 T - a_2 T^2 = 0.
\eqae
To zeroth order in the tachyon, the dilaton and metric backgrounds
are
\eq
\Phi_0 = 2\phi,\qquad G_{0\mu\nu} = \eta_{\mu\nu}.
\eqe
The tachyon $T$ is related to the massless scalar $S$ of previous sections by
\eq
T = e^{2\phi} S.
\eqe
The $\phi \to -\infty$ behavior of the tachyon background is given by the
linearized solution,
\eq
T_0 \sim (b_1 \phi + b_2) e^{2\phi}. \e{leadback}
\eqe
The constant $b_2$ is determined in terms of $b_1$ by the full nonlinear
tachyon interaction\cite{mecrit}, as we will see below.

Henceforth we work in conformal gauge, $ds^2 = - e^{2\rho} dx^+ dx^-$.
We again expand in powers of the incoming tachyon, $T = T_0 + T^{(1)}
+ T^{(2)} + T^{(3)} + \ldots$.  To the order we will be working, only the
first order correction to the gravitational and dilaton backgrounds enters.
Taking $\Phi = \Phi_0 + \delta$ and linearizing in $\delta$ and $\rho$,
the graviton-dilaton field equations to $O(T^2)$ can be written
\eqa
a_1 (\del_+ - 2)\Omega &=& - (\del_+ T)^2 + T^2  \nonumber\\
a_1 (\del_- + 2)\Omega &=& (\del_- T)^2 - T^2  \nonumber\\
2 a_1 \del_+ \del_- \delta &=& 2 a_1 \Omega + T^2,  \e{graveq}
\eqae
where $\Omega = 2(\del_- - \del_+) \delta + 4\rho$.
The tachyon equation is
\eqa
\del_+ \del_- S^{(1)} &=& - \frac{a_2}{2} T_0 S^{(1)} \nonumber\\
\del_+ \del_- S^{(2)} &=& -\frac{a_2}{4} e^{x^+ - x^-} (S^{(1)})^2
- \frac{a_2}{2} T_0 S^{(2)} \nonumber\\
\del_+ \del_- S^{3)} &=& \frac{1}{2} \Omega S^{(1)}
+ \del_+ \delta \del_- S^{(1)}
+ \del_- \delta \del_+ S^{(1)} \e{tacheq}\\
&&\qquad\qquad\quad  -\frac{a_2}{2} e^{x^+ - x^-}
S^{(1)} S^{(2)} - \frac{a_2}{2} T_0 S^{(3)}. \nonumber
\eqae

These are now solved using the retarded Green function $G(x^+,x^-)
= \theta(x^+) \theta(x^-)$, which satisfies $\del_+ \del_- G(x^+,x^-)
= \delta(x^+) \delta(x^-)$.  The initial condition is
\eq
S^{(1)}(t,\phi) \to S_+(x^-),\qquad S^{(2,3,\ldots)}(t,\phi) \to 0
\eqe
for $t \to -\infty$.  The leading behavior of the outgoing $S^{(1)}$ as
$x^+, x^- \to -\infty$ comes from the leading behavior~\ee{leadback} of the
background tachyon.  Integrating the first-order equation gives
\eq
S^{(1)}_- (x^+) \sim -\frac{a_2}{2}
\int_{-\infty}^{\infty} du^- \Bigl\{ b_1 (x^+ - u^-) + (b_2 - b_1) \Bigr\}
e^{x^+ - u^- }  S_+(u^-).
\eqe
This is the same as the matrix model result~\ee{first}, with
$ b_2 = b_1(-1 - 4\Gamma'(1) + \ln \mu) $ now determined, and $b_1
= -2\mu/a_2$.  The $\phi$ and $\mu$-dependence of the tachyon background is as
argued in ref.~\cite{mecrit}.

In the higher order equations~\ee{tacheq}, the effect of the background
tachyon is subleading as $x^+ \to -\infty$ (both the explicit terms, and the
implicit dependence through the graviton-dilaton back-reaction) and
so we ignore it.  The leading outgoing wave at second order is then
\eq
 S^{(2)}_-(x^+) \sim
-\frac{a_2}{4}
\int_{-\infty}^\infty du^-\, e^{x^+ -u^-} S_+^2(u^-),
\eqe
agreeing with the matrix model result~\ee{2otb} and
determining $a_2 = -2\sqrt{2}$ and $b_1 = \mu/\sqrt{2}$.
To third order we integrate the
graviton-dilaton equations~\ee{graveq} and then the tachyon equation to get
\eqa
S^{(3)}_-(x^+) &\sim&
\frac{1}{2} e^{2x^+} \int_{-\infty}^\infty du^-\,e^{-u^-}
S_{2+}(u^-) \int_{-\infty}^\infty dv^-\, e^{-v^-}  S_{1+}^2(v^-)
 \e{last}\\
&&\qquad - \frac{1}{4a_1} e^{2x^+} \int_{-\infty}^\infty du^-\,e^{-2u^-}
S_{2+}(u^-) \int_{-\infty}^\infty dv^-\, \dot S_{1+}^2(v^-). \nonumber
\eqae
Again this agrees, and fixes the final constant $a_1 = \frac{1}{2}$.

\section{Conclusions}

In a sense we have only worked out in coordinate space what is already known in
momentum space, that the difference between the trivial bulk S-matrix of the
matrix model and the nontrivial one of two-dimensional string theory is
the normalization of the vertex operators.  It is in coordinate space, however,
that the significance of the difference becomes clear: it is a non-local field
redefinition, which because of the simple kinematics in two dimensions can
convert the local matrix model interaction into the nonlocal interaction
from the gravitational and other higher fields
of string theory.

In particular one learns that the simplicity of the matrix model is rather
deceptive.  Consider the schematic representation in figure~2 of the
gravitational scattering (steps~I, II, and III are as defined below
eq.~\ee{trans}).  In step~I, the exponential pre-tail produced by
the convolution
of pulse~2 has an overlap with pulse~1.  In step~II the combined pulse reflects
off the end of the eigenvalue distribution, the ``wall.''  In step~III the
final
convolution produces an exponential pre-tail on the outgoing pulse, which is
the
bulk scattering of interest.  On the other hand, one believes that the actual
physical picture is that pulse~2 scatters off the gravitational field of
pulse~1
before it ever reaches the wall.  So the matrix model does not
reflect the qualitative physics of the scattering process.

One could extend our exercise to higher orders and so to higher string fields,
but it seems more efficient to try to work directly at the Lagrangian level.
The key seems to be to combine steps~I through~III
so as to write the exact solution in a way which correctly
represents the locality properties of the interaction.  The first step would
be the field redefinition~\ee{trans},
at least in the asymptotic free-field region.  The result will be a non-local
action, which presumably can be restored to a local form by additional
non-linear field redefinitions as well as the introduction of additional
non-dynamical fields (the string dilaton and metric, and higher).

We should emphasize that there is no local relation between the spacetime
metric
and the matrix model field.  Such a relation has occasionally been
proposed, but it is clear that it cannot exist because the gravitational field
at a given point must depend on the total energy {\it interior} to the point,
as
found in the scattering~\ee{finally}.

It is not immediately obvious how to produce a black hole from incoming
tachyons,
or to represent an eternal black hole in the matrix model.  The former question
in particular requires that we understand better the strongly nonlinear
solutions to the matrix model.

It has been proposed to identify the critical string tachyon with the
matrix model loop operator\cite{loops,mmreviews},
\eq
S'(t,\phi) = \int_{-\infty}^\infty dq \,\del_q \overline S(q,t)
e^{e^{\phi - q}}.
\eqe
Like the relation~\ee{trans} this is multiplicative in momentum space and a
convolution in position space,
but it is not of the same form and does not coincide
with the tachyon field that appears in the low energy Lagrangian.  We are not
sure of the relation, if any,
between our work and the studies of the macroscopic
loop operators.  We note in passing that our $S$ satisfies a linearized
equation with the tachyon background~\ee{leadback} having a linear term,
whereas
the loop operator $S'$ satifies a linearized equation with no linear term in
the background.

Other nonlocal transformations of the tachyon field have played a role in the
matrix model black hole proposals of refs.~\cite{Das,DMW,JevYon}.  We again are
not sure of any relation between this work and ours, but we should note that we
are puzzled by the proposal \cite{JevYon} that processes with odd numbers of
tachyons should vanish in the black hole background.  There is no sign of any
$Z_2$ symmetry in the effective spacetime action~\ee{stact}, and the $2 \to 1$
process that we have discussed should still occur in the region exterior to the
horizon.  We should also note ref.~\cite{Russo}, which discusses dynamical
processes in the matrix model.  This work does not include a nonlocal
transformation of the tachyon, and so proposes a local relation between the
metric and the matrix model fields.

In summary, the existence of the exact matrix model solutions to
low-dimensional string theories ought to be a useful tool for understanding
string physics in spacetime.  The relation between the matrix models
and the string has been a subject of some confusion.  We hope that our work
helps to clarify this subject.

\centerline{\bf Acknowledgements}

We would like to thank S. Chaudhuri, M. Douglas, M. Stone, and A. Strominger
for
discussions.  This work was supported in part by
National Science Foundation grants PHY89-04035 and PHY91-16964.

\vfill

\pagebreak

\centerline{\bf Figure Captions}

\begin{itemize}

\item[1.] Successive pulses moving in the $\phi$-$t$ plane.  Gravitational
field
of pulse~1 (dotted) should cause part of pulse~2 to backscatter, producing an
outgoing wave (dashed) which precedes the main reflection from the `wall.'

\item[2.] How the matrix model represents the process of figure~1.  The initial
wavefunction renormalization (I) produces a tail on pulse~2 which overlaps
pulse~1; the combined pulse reflects from the wall (II); and the final
renormalization (III) produces the outgoing wave.

\end{itemize}

\pagebreak

\end{document}